\documentclass[a4paper,twocolumn,pra]{revtex4-1}
\usepackage{graphicx,amsmath}
\newcommand{\beq}{\begin{eqnarray*}}
\newcommand{\eeq}{\end{eqnarray*}}
\newcommand{\beqn}{\begin{eqnarray}}
\newcommand{\eeqn}{\end{eqnarray}}
\newcommand{\abs}[1]{\ensuremath{\lvert#1\rvert}}

\newcommand{\op}[1]{{\hat{#1}}}
\newcommand{\bket}[1]{\mbox{$|{#1}\rangle$}}

\newcommand{\ip}[2]{\mbox{$\langle{#1}|{#2}\rangle$} }
\begin{document}


\title{Aspects of Complementarity and Uncertainty}
\thanks{Published: \bf Int. J. Quant. Inf. 14(3), 1640031 (2016) }

\author{Radhika Vathsan}

\address{Department of Physics, BITS Pilani K K Birla Goa Campus\\
Zuarinagar, Goa, India.\\
radhika@goa.bits-pilani.ac.in}

\author{Tabish Qureshi}

\address{Center for Theoretical Physics,
Jamia Millia Islamia\\ New Delhi, India.\\
tabish@ctp-jamia.res.in}


\begin{abstract}
The two-slit experiment with quantum particles provides many insights into the behaviour of quantum mechanics, including Bohr's complementarity principle. Here we analyze Einstein's recoiling slit version of the experiment and show how the inevitable
entanglement between the particle and the recoiling slit as a which-way detector is responsible for complementarity. We derive the Englert-Greenberger-Yasin duality from this entanglement, which can also be thought of as a consequence of sum-uncertainty relations between certain complementary observables of the recoiling slit. Thus, entanglement is an integral part of the which-way detection process, and so is uncertainty, though in a completely different way from that envisaged by Bohr and Einstein.

\end{abstract}
\maketitle

\section{Introduction}
The two-slit experiment that we first encounter in introductions to Quantum Mechanics, is one of the 
most useful tools to explore foundational issues in the subject. It is a striking illustration of the principle of complementarity of Bohr \cite{bohr}, also sometimes referred to as wave-particle duality. The two-slit experiment captures the essence of quantum theory in such
a fundamental way that Feynman \cite{feynman} went to the extent of stating that it is a
phenomenon ``which has in it the heart of quantum mechanics; in reality it
contains the {\em only} mystery" of the theory. 

In his 1928 Como lecture, Bohr avers that in describing the results of quantum mechanical experiments, certain physical concepts are complementary. If two concepts are
complementary, an experiment that clearly illustrates one concept will
obscure the other complementary one. In a two-slit experiment with quantum particles, the {\em interference} pattern
embodies their wave nature, while the detection of {\em which slit} the particle passed through is the complementary particle nature. These two, said Bohr, cannot both be seen in the same run of the experiment.

\begin{figure}[h]
\begin{tabular}{cc}
 \includegraphics[width=0.5\columnwidth]{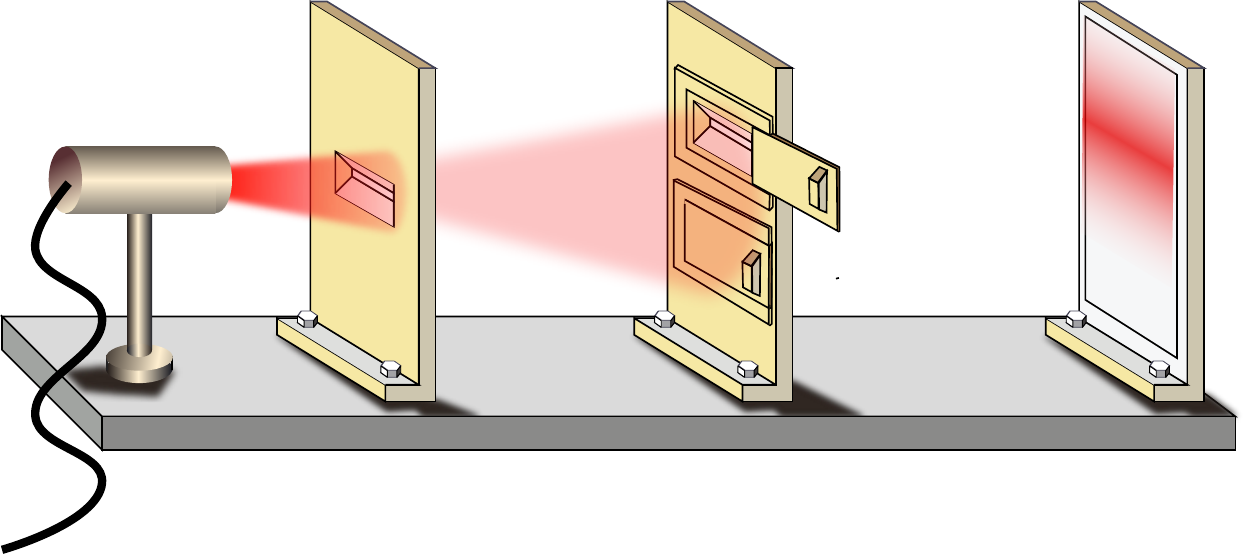} &
  \includegraphics[width=0.5\columnwidth]{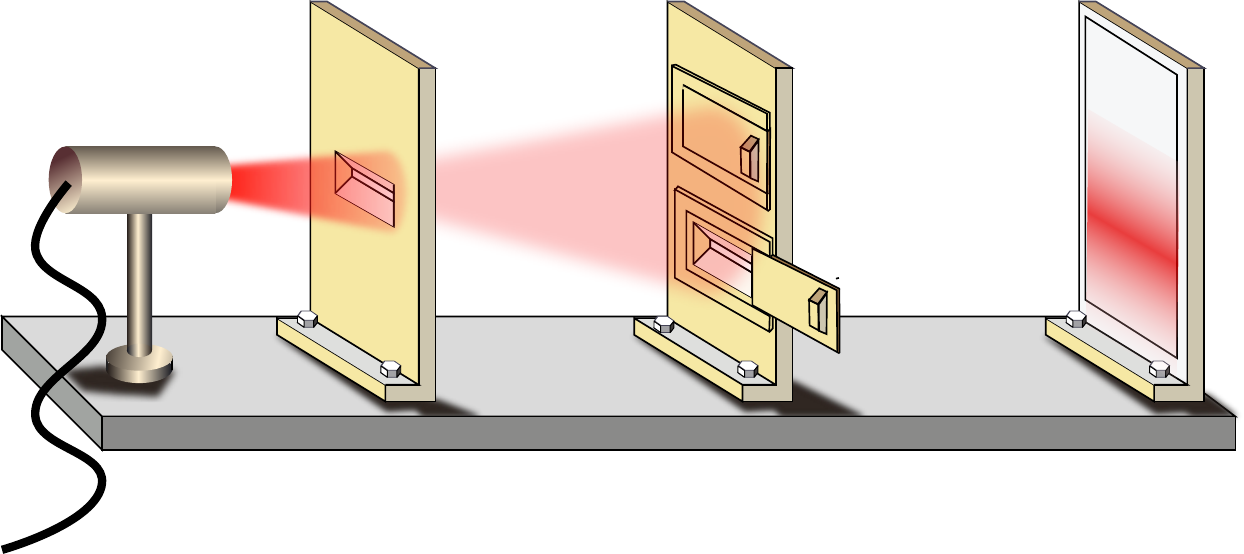} \\
  (a) & (b) \\
 \includegraphics[width=0.5\columnwidth]{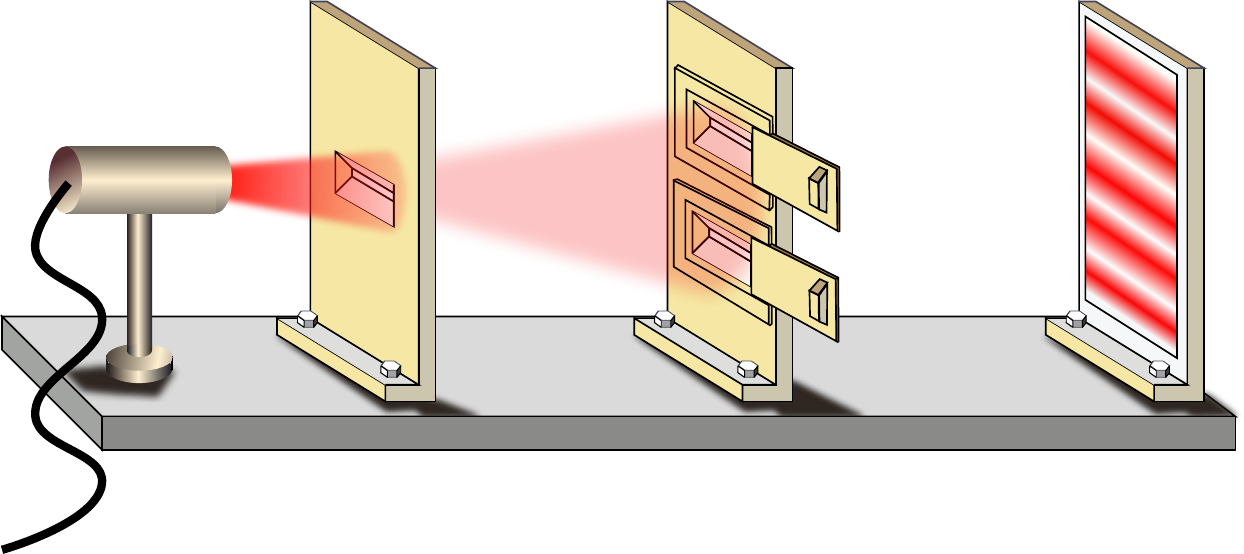} &
 \includegraphics[width=0.5\columnwidth]{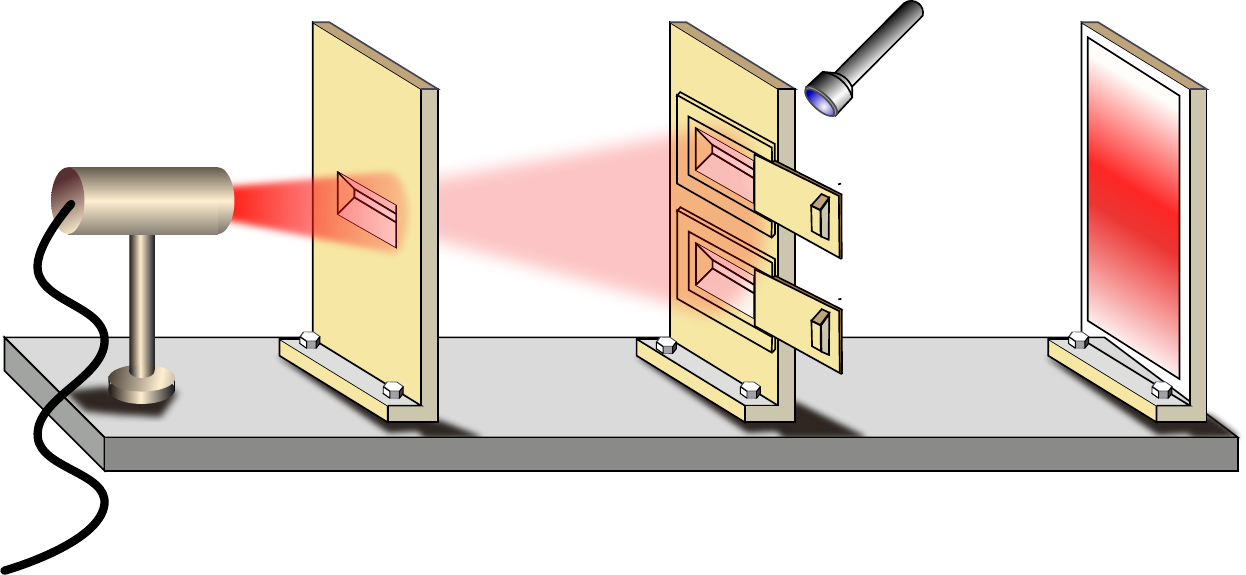} \\
 (c) &(d)\\
 \end{tabular}
 \caption{The traditional double-slit experiment with quantum particles.  When either slit is closed, there is no interference but when both slits are open, the screen shows typical maxima and minima of intensity characteristic of wave-like behaviour. (d) shows that which-way detection destroys the interference pattern.}
 \label{fig:double-slit}
\end{figure}

Bohr's explanation for this complementarity relied on the Heisenberg indeterminacy principle, according to which the process of measuring which-way information would disturb the particle to just the right extent to wash out the interference fringes. The key feature in his argument was that the process of measurement that yielded the which-way information should be treated carefully, and the measuring instrument should also be regarded as a quantum mechanical system. To emphasize this point, Bohr drew semi-realistic images of the experimental apparatus (see fig \ref{fig:double-slit}) with almost exaggerated emphasis on the nuts and bolts so that it was clear that when a quantum system was being measured, the measuring apparatus must be treated on the same footing.

Einstein's discomfort with quantum indeterminacy (``God does not play dice"), in his famous arguments with Bohr, sought to explain away the consequences predicted by indeterminacy using classical concepts 
of energy and momentum conservation. In the context of the double slit experiment, he proposed a modification to indirectly obtain the which-way information, which, he claimed, would not affect the wave nature, thus allowing us to access both the ``complementary" natures of the quantum particle in the same experiment.

\section{The Recoiling Slit Experiment}
Einstein modified Bohr's picture to allow the source slit to move. He
suggested that it be mounted in delicate springs (or rollers, as in
later versions of the same experiment), equipping it with a degree of
freedom along the $x$-axis (see Fig. 2). The idea was that when the quantum particle
``diffracted'' from this source slit to move towards the upper or lower
slit on the next screen, it is deflected from its original direction along
the $y$ axis. A particle emerging from the upper slit for instance must
have  acquired a (small) component of momentum along the $x$-axis when
it left the source screen. This would therefore cause the source screen
to recoil with an equal opposite momentum for momentum conservation
to hold. Measuring this recoil momentum would thus provide a means
of detecting which-way without affecting the further propagation of
the particle on its way towards the interference pattern on the final
screen. Einstein triumphantly concluded that the sacrosanct principle of
momentum conservation would thus allow an experimental determination  of
which-way in the same  experiment that showed up the interference pattern.

\begin{figure}[h!]
 \includegraphics[width=1.0\columnwidth]{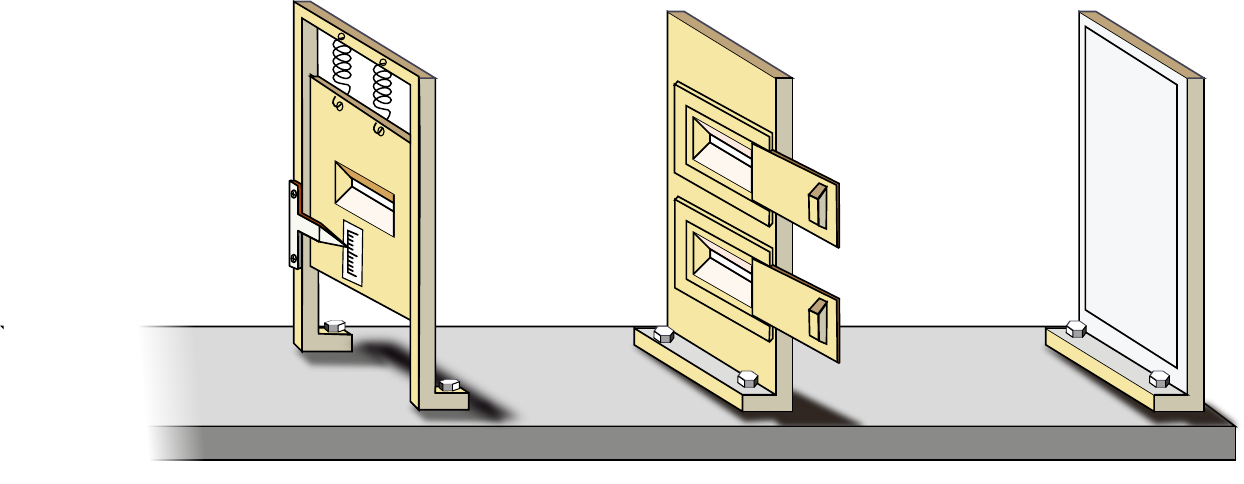}
 \caption{Einstein's recoiling slit setup}
\end{figure}

Bohr's rebuttal to Einstein relied again on a careful treatment of the measuring apparatus, now the recoiling slit, as a quantum object. His argument\cite{recoil} was that in order to obtain reliable which-way information, the momentum of recoil must be measured to a certain degree of accuracy. This meant the initial momentum of the source slit must be known to the same accuracy. The source slit being a quantum object, this meant that its initial {\it position} (along the $x$-axis)
 must be uncertain in a way as to satisfy the uncertainty principle. This uncertainty, claimed Bohr, was just sufficient to wash out the interference pattern on the final screen! Remember that the interference pattern on the screen is the cumulative result of hits by various individual particles, and the spread in the origin of the coordinate system results in each particle being part of an interference pattern slightly shifted with respect to the others, so that the net outcome is the blurring out of the maxima and minima.

 Here is a quick, back-of-the-envelope calculation that convinces us that Bohr is right.
 
 \begin{figure}[h]
 \center\includegraphics[width=1.0\columnwidth]{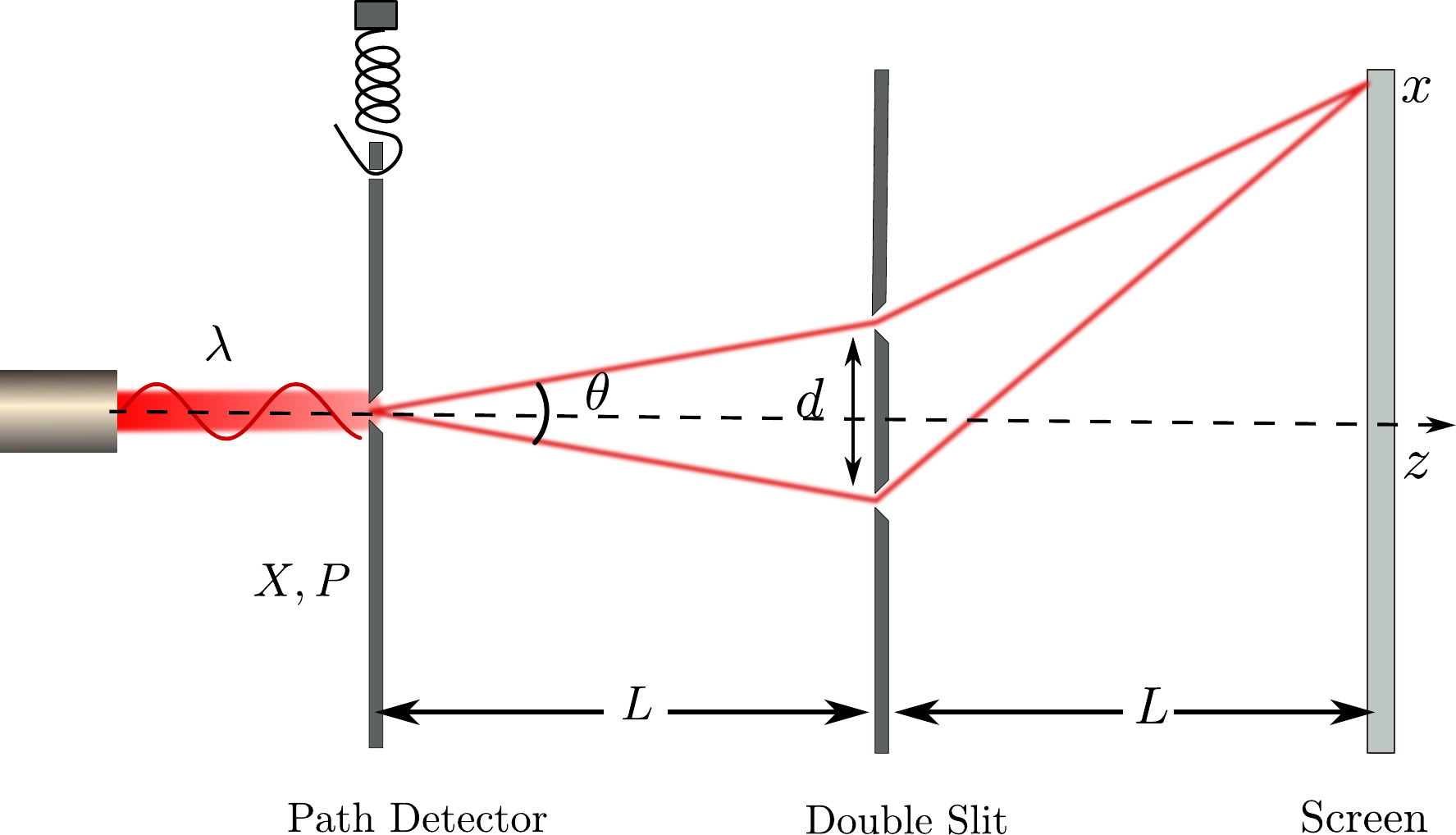}
 \caption{Geometry of the recoiling slit experiment}
 \label{fig:recoilslit}
\end{figure}
Refer to Figure~(\ref{fig:recoilslit}) for the parameters involved in the experiment, for particles of average momentum $p$ and d'Broglie wavelength $\lambda$.
The spread in $x$-momentum between particles passing through the upper and the lower slit  is
 \[\Delta p_x = 2p\sin({\theta/2}) \approx p\theta =
\frac{h}{\lambda}\theta = \frac{h}{\lambda} \frac{d}{ L}.\]  
This is the limit on  accuracy of measuring recoil momentum.
Invoking the uncertainty principle, the minimum indeterminacy in the  position of  the source slit is
$\displaystyle \Delta x = \frac{\hbar}{ 2\Delta} p_x = \frac{\lambda L}{ 4\pi d}$.
This is to be regarded as the  uncertainty in the position of a fringe on the final screen.
Now the fringe separation by Young's formula is $\delta x =\frac{\lambda L }{d}$, the same order of magnitude as the uncertainty in the position. This is why the interference pattern is lost.

The subtle ideas evoked by this experiment have triggered a lot of research in the subject. Among the earlier work,
Wooters and Zurek\cite{wooters} carried out a quantitative analysis of Bohr's argument,
assuming the recoiling slit to be constrained by a harmonic oscillator
potential. There have subsequently been several experimental vindications
of Bohr \cite{haroche}: complementarity is indeed true! The first such experiment was
reported by Utter and Feagin\cite{utter}, who used a trapped ion in
place of the recoiling slit.

\section{Theoretical Analysis}

\subsection{Which-way information and entanglement}

Now Bohr's argument seems to favour complementarity as a restatement of the uncertainty principle: the latter enforces the former. However, a crucial aspect of the measurement process: namely the {\em entanglement} between the measuring device and the system, was not part of Bohr's argument\cite{note1}.
This picture of measurement  arose  later, in von-Neumann's quantum measurement model. According to von Neumann\cite{neumann}, measurement of a property of a quantum system can be regarded as a two-part process. First, the measuring device and the quantum system must get coupled by a unitary evolution, causing their  states to become entangled. 
Suppose the system is initially in a superposition state
\[\bket{\psi_0} = \sum_{i=1}^n c_i\bket{u_i} \]
expressed in some basis $\{\bket{u_i} \} $. The measuring device may be thought of as being in some initial state $\bket{d_0}$. von Neumann's first process is an interaction between the system and detector leading to entanglement of the detector and system states: 
\beqn
\bket{\Psi_0} = \bket{d_0} \otimes \sum_{i=1}^n c_i \bket{u_i}
\xrightarrow{\textrm{Unitary evolution}} \sum_{i=1}^n c_i \bket{d_i}\bket{u_i}.
\label{process1} 
\eeqn
Next, the measuring device is subjected to a non-unitary process that causes it to collapse to one particular state, picking out one of several possible outcomes for the measurement. For example, if the device pointer ends up in a state $\bket{d_k}$ after this process, we have the collapse of the system-detector state
\beqn
\sum_{i=1}^n c_i\bket{d_i}\bket{ \psi_i}
\xrightarrow{\textrm{Process~2}}\bket{d_k}\bket{\psi_k}
\label{process2}
\eeqn
The second process, which is the heart of the so-called {\em quantum measurement problem}, does not concern us here. We will examine the which-way detection with regard to the first  process, and show how this establishment  of correlations between the system and the recoiling slit is alone sufficient to enforce complementarity.

Applying this to the recoiling slit experiment, the  $x$-states of the particle may be regarded as  states corresponding to the positions of the  two slits. Let's call them $\bket{\psi_1}$ and $\bket{\psi_2} $. Correspondingly, the recoiling slit has momentum states $\bket{d_1}$ and $\bket{d_2}$. It is in principle possible to find an interaction between the particles and the detector such that $\bket{\psi_1}$ and $\bket{\psi_2}$  are unaffected, but the detector states get entangled with them:
\beqn
\bket{\Psi} &=& c_1 \bket{d_1}\bket{\psi_1} + c_2 \bket{d_2}\bket{\psi_2}.
\eeqn
This alone is sufficient to wash out the interference pattern on the final screen! Suppose on reaching the screen at some position $x$, the particle states evolve  to $\bket{\psi_1(x,t)}$ and $\bket{\psi_2 (x,t)}$. We assume without loss of generality that the detector states do not evolve in this time. So the combined detector-particle state is now
\beqn
\bket{\Psi(x,t)} &=& c_1 \bket{d_1}\bket{\psi_1(x,t)} + c_2 \bket{d_2}\bket{\psi_2(x,t)}.
\eeqn
The probability of detecting the particle at this location on the screen is therefore given by
\beqn
\abs{\Psi(x,t)}^2 &=& \abs{c_1}^2\abs{d_1}^2\abs{\psi_1(x,t)}^2
+ \abs{c_2}^2\abs{ d_2}^2\abs{\psi_2(x,t)}^2 \nonumber\\
&+& c_1^*c_2\ip{ d_1}{d_2}\ip{\psi_1(x,t)}{\psi_2(x,t)}
+ c_2^*c_1\ip{ d_2}{d_1}\ip{\psi_2(x,t)}{\psi_1(x,t)},\nonumber\\
\eeqn
where we have used the expedient $\abs{\psi}^2 = \ip{\psi}{\psi}$. The last two terms here denote interference.
Now if the detector states are distinguishable, then $\ip{d_1}{d_2}=0$, implying that the interference terms vanish!
The mere fact that the detector {\em carries which-way information} is sufficient to wash out interference effects: there is no need to invoke position-momentum uncertainty of the recoiling slit. It is important to realize that this happens regardless of the method used to distinguish the which-way information. Any variant of this experiment, such as that proposed by Scully, Englert and Walther\cite{scully} will also yield the same result.

\subsection{Path-distinguishability and Interference: Gaussian wave-packet model}
We normally assume that the detector  states are distinct. But this need not be true in general. The more interesting cases are between the two extremes of perfect distinguishability and no distinguishability. We could have chosen to look at detector states that are not fully orthogonal. This would mean that the paths taken by the particle are not perfectly distinguishable. The implications of this for the interference fringes was analyzed by Englert\cite{englert}, where   a duality between fringe-visibility and path distinguishability was derived. This duality was first analyzed in an experimental context by Greenberger et al\cite{greenberger}, and subsequently discussed 
theoretically by Jaeger et al\cite{jaeger}.

We first define a quantitative measure of the {\em distinguishability} of the paths, the probability with which the path taken by the particle is correctly given by looking at the detector states $\bket{d_1}$ and $\bket{d_2}$. The definition  depends on the knowledge of which-path that is obtainable from a given which-way detector, and could also depend on the state preparation of the system. For our purposes, the definition
\beqn
\mathcal{D} &=& \sqrt{1 -\abs{\ip{d_1}{d_2}}^2}
\label{dist}
\eeqn
is sufficient.
Let's see how this is justified. In order to obtain the which-way information stored in the detector, we need to measure a  suitable observable $\op{W}$ of the detector, with distinct eigenvalues $w_1$ and $w_2$ and corresponding eigenstates $\bket{w_1}$ and $\bket{w_2}$. 
Suppose the detector states are expressed in this basis as
\beqn
\bket{d_1}  &=&  \bket{w_1} \nonumber \\
\bket{d_2} &=& \alpha_w \bket{w_1} + \alpha_r \bket{w_2}
\eeqn 
They are explicitly non-orthogonal. If the particle passes through slit 1, detector state $\bket{d_1}$ gives us the correct information. But if it passes through slit 2, the state $\bket{d_2}$ has correct information with probability $\abs{\alpha_r}^2$.
On measuring $\op{W}$ if we obtain $w_2$, the probability of getting the right answer is
\beq
\abs{\alpha_r }^2 &=& |\langle w_2 | d_2 \rangle |^2 \\
              &=& 1 - |\langle w_1 | d_2 \rangle |^2\\
              &=& 1 - |\langle d_1 | d_2 \rangle |^2 \\
              &=& \mathcal{D}^2
              \eeq
So $\mathcal{D}$ is the probability amplitude of correctly distinguishing the two paths.

If the detector states are orthogonal then $\mathcal{D} =1 $ and if they are the same then  $\mathcal{D}=0$.
Further refinement of the notion of distinguishability is discussed, for example, by Englert\cite{englert}.
We now model the state of the incident particle traveling along the $z$-axis by  a Gaussian wave-function centered on the source slit, with width $\epsilon$:
\beqn
g(x) &=& \frac{1}{\sqrt[4]{8\pi \epsilon^2}} \exp{\left(-\frac{x^2}{ 4\epsilon^2} \right)}.
\eeqn
We are not explicitly considering shape of the wave-function along the $y$ direction: it is not relevant to the discussion below.
At $t=0$, the particle strikes the double-slit  and emerges, after interacting with the detector by a process like  that of Eq.~(\ref{process1}), with correlated wave-function
\beqn
\Psi(x,0) &=&  A \left[\bket{d_1} g(x-d/2)
+ \bket{d_2}  g(x+d/2)\right],\\
&& A =
\frac{1}{\sqrt[4]{8\pi \epsilon^2}}\nonumber 
\eeqn
Here, we will not explicitly consider the dynamics along the forward direction. We will  assume that the wave-packets are moving in $z$-direction with an average momentum $p_0=h/\lambda_d$, where 
$\lambda_d$ is the d'Broglie wavelength of the particle. The distance $z$ traveled  in a time $t$  given by $z = \frac{h}{m\lambda_d}t$. This can be rewritten as
$\hbar t/m = \lambda_d z/2\pi$. Each Gaussian wave-packet then spread to a  new Gaussian defined by
\beqn
g(x,t) &=& A_t \exp \left( -\frac{x^2}{ 4\epsilon^2+2i\hbar t/m} \right),  \\
\textrm{ where }A_t &=& \frac{1}{\sqrt{2}}\left[\sqrt{2\pi}(\epsilon+i\hbar t/2m\epsilon)\right]^{-1/2}. \nonumber
\eeqn

We also assume that the detector states do not evolve after this interaction, so that the combined state of the particle and detector after time $t$ evolves to
\beqn
\Psi(x,t) &=& A_t \left( \bket{d_1} b(x-d/2,t)
+ \bket{d_2} b(x+d/2,t)\right).
\label{entstatet}
\eeqn

The probability of finding the particle at position $x$ on the screen is
given by
\beqn
\abs{\Psi(x,t)}^2 &=& \abs{A_t}^2 \left(e^{-\frac{(x-d/2)^2}{ 2\sigma_t^2}}
          + e^{-\frac{(x+d/2)^2}{ 2\sigma_t^2}}\right)\nonumber\\
&+& \abs{A_t}^2 \left(\ip{ d_1}{d_2} e^{-\frac{x^2+d^2/4}{2\sigma_t^2}} 
      e^{\frac{ixd\hbar t/2m\epsilon^2}{2\sigma_t^2}} \right.\nonumber\\
&& +\left. \ip{d_2}{d_1} e^{-\frac{x^2+d^2/4}{2\sigma_t^2}}
     e^{-\frac{ixd\hbar t/2m\epsilon^2}{2\sigma_t^2}}\right) ,
\eeqn
where $\sigma_t^2 = \epsilon^2 + (\hbar t/2m\epsilon)^2$. Writing
$\ip{ d_2}{d_1}$ as $\abs{\ip{d_2}{d_1}}e^{i\theta}$, and
putting $\hbar t/m = \lambda_dL/2\pi$, the above can be simplified to
\begin{eqnarray}
\abs{\Psi(x,t)}^2 &=& \abs{A_t}^2 e^{-\frac{x^2+d^2/4}{2\sigma_t^2}}
\cosh(x d /2\sigma_t^2) \nonumber\\
&&\times\left(1 + \abs{\ip{ d_1}{d_2}} 
\frac{\cos\left({\frac{xd\lambda_dL/2\pi}{4\epsilon^4+(\lambda_dL/2\pi)^2}+\theta}\right)}{
\cosh(x d /2\sigma_t^2)} \right)\nonumber\\
\label{pattern}
\end{eqnarray}
Eq.~(\ref{pattern}) represents an interference pattern with a fringe width
given by 
\begin{equation}
w = 2\pi\left(\frac{(\lambda_dL/2\pi)^2 + 4\epsilon^4}{\lambda_ddL/2\pi}\right)
= \frac{\lambda_dL}{d} + \frac{16\pi^2\epsilon^4}{\lambda_ddL}.
\end{equation}
For $\epsilon^2 \ll \lambda_dL$ we get the familiar Young's double-slit
formula $w \approx \lambda_dL/d$.

The visibility of the interference pattern is conventionally defined as the contrast in intensities of neighbouring fringes
\begin{equation}
{\mathcal V} = \frac{I_{\rm{max}} - I_{\rm{min}}}{ I_{\rm{max}} + I_{\rm{min}} } ,
\end{equation}
where $I_{\rm{max}}$ and $I_{\rm{min}}$ represent the maximum and minimum intensity
in neighbouring fringes. The fringe visibility actually depends on the coherence of the 
``waves'' when they arrive at the screen, and  on the geometry of the setup, such as the width of the slits and their separation. For example, if
the width of the slits is very large, the fringes may not be visible at all.
Maxima(minima) of Eq.(\ref{pattern}) will occur at points where the 
value of cosine is $1$($-1$). The visibility can then be 
written down as
\begin{equation}
{\mathcal V} = \frac{\abs{\ip{d_1}{d_2}}}{ \cosh(x d /2\sigma_t^2)}.
\end{equation}
Since $\cosh(y) \ge 1$, we get
\begin{equation}
{\mathcal V} \le \abs{\ip{ d_1}{d_2}}.
\label{visiblity1}
\end{equation}
Using Eq.~(\ref{dist}) the above equation yields the important Englert-Greenberger-Yasin duality
relation
\begin{equation}
{\mathcal V}^2 + {\mathcal D}^2 \le 1.
\label{duality}
\end{equation}
Eq.~(\ref{duality}) can be considered as a quantitative statement of
Bohr's complementarity principle. It sets a bound on the which-path 
distinguishability and the visibility of interference that one can
obtain in a single experiment. In particular, if the distinguishability is perfect ($1$) then the visibility is zero, and if the visibility is perfect then distinguishability is zero. Notice that this has nothing to do with uncertainty between  any pair of complementary variables: it is purely a consequence of the quantum nature of the detector and of the entanglement between the detector and particle states.

\subsection{Uncertainty and duality}
Notwithstanding what we just concluded in the previous section, there is also another
view prevalent in the literature which holds that the process of which-way
detection introduces certain uncontrollable phases to the state of
the particle, which leads to loss of interference\cite{tan,storey}.
The uncertainty relation is believed to play a role in the latter.
Whether complementarity arises
out of correlations between the particle and a which-path detector or
from the uncertainty principle, has been a subject of some controversy
\cite{storey,englert2,wiseman,barad}.
Linked to this controversy is also the question whether the particle
receives any momentum kick from the recoiling slit, affecting its interference
pattern\cite{wiseman2,durr,unni}.
There have been various approaches to connect complementarity to 
uncertainty relations \cite{bjork,marzlin,huang,bosyk,shilladay}. 

The duality relation as we have seen, arises due to the obtaining of which-way information by observing the detector states. The detector (the recoiling slit in this case), acquires one of two momentum states when the particle passes through the double-slits. We measure the state of the detector through observation of some dichotomic observable, say 
$\op{P}$ with eigenvalues $\pm 1$ and corresponding eigenstates $\bket{p_1}, \bket{p_2}$. In general, the detector states that get correlated with the particle states are $\bket{d_1}, \bket{d_2}$, which are normalized but not necessarily orthogonal. They can be expressed in the basis of $\op{P}$ eigenstates, without loss of generality, as
\beqn 
\bket{d_1} &=& c_1 \bket{p_1} + c_2 \bket{p_2} \\
\bket{d_2} &=& c_2^*\bket{p_1} + c_1^*\bket{p_2}.
\eeqn
When for instance $|c_1|=1,~c_2=0 $ (or the other way round), these states carry full which-way information, and when
$|c_1|=|c_2|=1/\sqrt{2}$, they carry no which-way information. This form for the detector states covers all cases of mutual overlap.

Now looking at the  measurement statistics  of $\op{P}$ in either of $\bket{d_1}, \bket{d_2}$, the variance is
\beqn
\Delta{P}^2 = \langle \op{P}^2\rangle - \langle \op{P} \rangle^2
                  = 4 \abs{c_1}^2 \abs{c_2}^2.
\eeqn
Also since $ \abs{\ip{d_1}{d_2}}^2 = 4 \abs{c_1}^2 \abs{c_2}^2 $, distinguishability as defined by Eq.~(\ref{dist}) is given by
\beqn
\mathcal{D}^2 = 1 - \Delta {P}^2.
\label{dist2}
\eeqn
Thus for perfect distinguishability of path, the variance in $\op{P}$ should be zero.

Now the interference pattern on the final screen, built by successive registering of individual particles hitting the screen, can also be explained by considering the phase shifts acquired by each particle on passing through the double slit. We can think of this phase shift as accruing due to the interaction between the particle and the which-way detector\cite{unni}. This was the approach taken by Bohr. We can use this approach  link the fringe visibility to uncertainty in some observable associated with the detector.

 Prior to measuring the detector observable $\op{P}$, consider the system-detector entangled state  of the form
\beqn
\bket{\Psi} = \bket{\psi_1} \bket{p_1} + \bket{\psi_2}\bket{p_2}.
\eeqn
But now suppose we  change the basis for the detector states to
\beqn
\bket{q_1} = \frac{1}{\sqrt{2}}[\bket{p_1} + \bket{p_2} ],  ~~~
\bket{q_2} = \frac{1}{\sqrt{2}}[\bket{p_1} - \bket{p_2} ].
\eeqn
Measurement of the detector states in this basis corresponds to the measurement of a different dichotomic observable $\op{Q}$ with eigenstates $\bket{q_1}, \bket{q_2}$.
The particle states entangled with these detector states can be expressed as
\beqn
\bket{\Psi} = \frac{a_1}{\sqrt{2}}[\bket{\psi_1} + \bket{\psi_2}]\bket{q_1}
            + \frac{a_2}{\sqrt{2}}[\bket{\psi_1} - \bket{\psi_2}]\bket{q_2}
\label{entstate2}
\eeqn
where $\abs{a_1}^2 + \abs{a_2}^2 =1$.
Suppose we look at the symmetric case $a_1=a_2$, something interesting is observed regarding this basis.
If we correlate the measurements of $\op{Q}$ with the detections on the screen, then there is an interference pattern, but if we do not, then the pattern vanishes. In the latter case, even though the which-way detector is entangled with the particle states, and we are measuring the states of the which-way detector, there is no interference: in this basis, there is effectively no collection of which-way information. This is easy to see from the form of the $\op{Q}$ eigenstates: they are equal superpositions of the $\op{P}$ eigenstates: they are states in which slit 1 and slit 2 are equally probable: and therefore which-path information is not obtained in this basis. This is an example of the quantum eraser \cite{eraser}.

One way to understand this is the following: correlating the $Q=+1$ particles with their positions on the screen shows an interference pattern, and so does the same for the $Q=-1$ particles, but these two patterns are shifted with respect to each other: one set is the ``antifringe'' of the other: so that the two together cancel each other (see Fig. 4).
\begin{figure}[h]
 \center{\includegraphics[width=1.0\columnwidth]{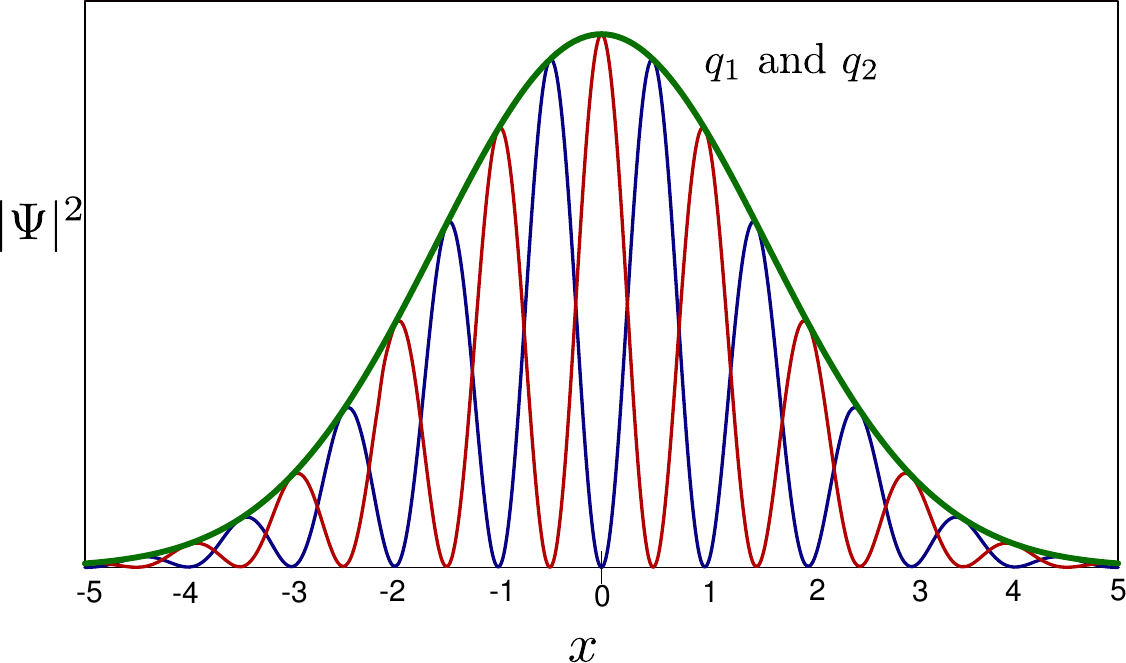}}
 \caption{Correlation of detected particles corresponding to $q_1$ and $q_2$ gives two complementary interference patterns. The red curve corresponds to $q_1$ and the blue to $q_2$. The envelop of these two corresponds to not correlating the detections to the slit observables, and shows no interference pattern}
\end{figure}

Let's see how the measurements in the $\op{Q}$-basis link to fringe visibility.
The probability of detecting a particle at $x$ on the screen is
\beqn
\abs{\Psi(x)}^2 &=& \frac{1}{2} \left(\abs{\psi_1(x)}^2 + \abs{\psi_2(x)} \right)\nonumber\\
&&+ \frac{1}{2} (\abs{a_1}^2 - \abs{a_2}^2) \left(\ip{\psi_1}{\psi_2} + \ip{\psi_2}{\psi_1} \right).
\eeqn
From here it is straightforward to see that the visibility is limited by
\beqn
\mathcal{V}^2 &\le& (\abs{a_1}^2 - \abs{a_2}^2)^2.
\eeqn
Now measurement statistics on the state  in Eq.~(\ref{entstate2}) gives the uncertainty in the values of $\op{Q}$ as \cite{ndh-tq}
\beqn
\Delta Q^2 = 1 - (\abs{a_1}^2 - \abs{a_2}^2)^2
\eeqn
which can be inserted into the visibility inequality to give
\beqn
\mathcal{V}^2 &\le& 1- \Delta Q^2.
\eeqn
Combining this with Eq.~(\ref{dist2})
we get
\beqn
\mathcal{D}^2 + \mathcal{V}^2 &\le& 2- (\Delta P^2 + \Delta Q^2).
\label{duality2} 
\eeqn

However, the $\op{P}$ operator is complementary to $\op{Q}$. It would help us to remember that dichotomic observables that are complementary can be represented by two of the Pauli matrices. For instance, $\op{P}$ can be represented by $\sigma_3$ while $\op{Q}$ by $\sigma_2$.  Now any two components of the spin triad satisfy the sum uncertainty relation\cite{hofmann}
\beqn
\Delta\sigma_2^2 + \Delta\sigma_3^2 \ge 1,
\eeqn
which neatly implies the Englert-Greenberger-Yasin duality from Eq.~(\ref{duality2}):
\beqn
{\mathcal D}^2 + {\mathcal V}^2 \le 1.
\label{duality3}
\eeqn

Similar conclusions are also reached in earlier work by Duerr et al \cite{duerr}.
Thus even though the complementarity relations do not seem to be related to any position-momentum uncertainties for the recoiling slit, there does seem to be a connection to sum-uncertainty relations for a pair of complementary observables for the detector.  It should also be clear that position-momentum uncertainties would not apply to all schemes of which-way detection. Thus it would seem that complementarity is closely related to correlations between the system and the which-way detector, and not to position-momentum uncertainty relations.

\section*{Acknowledgments}

We thank the organizers of the International Program on Quantum Information, 2014, for providing an opportunity to interact with many workers in the area. We are indebted to Prof B-G Englert for providing insightful suggestions. We would like to thank Prof A R Usha Devi and  H. S. Karthik for interesting discussions.

\vspace*{-6pt}   

\end{document}